\documentclass[french]{gretsi}

\usepackage[utf8]{inputenc}

\usepackage[T1]{fontenc}

\usepackage{amsmath, amssymb, amsfonts}

\usepackage{cite} 
\usepackage{algorithmic} 
\usepackage{graphicx} 
\usepackage{textcomp} 
\usepackage{xcolor} 
\usepackage{hyperref} 
\usepackage{pgfplots}
\usepackage{subfig}

\pgfplotsset{compat=1.18}

\newcommand{\cc}{\mathcal{C}} 
\newcommand{\gggg}{\mathcal{G}} 
\newcommand{\vvv}{\mathcal{V}} 
\newcommand{\eee}{\mathcal{E}} 
\newcommand{\cb}{\mathbf{C}} 
\newcommand{\mathbfit}[1]{\textbf{\textit{#1}}} 
\newcommand{\mat}[1]{\mathbf{#1}}

\usepackage{todonotes}

\begin{document}


\titre{Génération de Matrices de Corrélation avec des Structures de Graphe par Optimisation Convexe}

\auteurs{
  \auteur{Ali}{Fahkar}{ali.fahkar@univ-grenoble-alpes.fr}{1}
  \auteur{Kévin}{Polisano}{kevin.polisano@univ-grenoble-alpes.fr}{1}
  \auteur{Irène}{Gannaz}{irene.gannaz@grenoble-inp.fr}{2}
  \auteur{Sophie}{Achard}{sophie.achard@univ-grenoble-alpes.fr}{1}
}

\affils{
  \affil{1}{Univ. Grenoble Alpes, CNRS, Grenoble INP, Inria, LJK, F-38000 Grenoble, France 
  }
  \affil{2}{Univ. Grenoble Alpes, CNRS, Grenoble INP, G-SCOP, 38000 Grenoble, France
  }
}


\resume{Ce travail porte sur la génération de matrices de corrélation théoriques présentant des motifs de parcimonie spécifiques, associés à des structures de graphe. Nous présentons une nouvelle approche basée sur l'optimisation convexe, offrant une plus grande flexibilité par rapport aux techniques existantes, notamment en contrôlant la moyenne de la distribution des entrées dans les matrices de corrélation générées. Cela permet de produire des matrices de corrélation représentant plus fidèlement des données réalistes et pouvant être utilisées pour l'évaluation comparative des méthodes statistiques d'inférence de graphes.}

\abstract{This work deals with the generation of theoretical correlation matrices with specific sparsity patterns, associated to graph structures. We present a novel approach based on convex optimization, offering greater flexibility compared to existing techniques, notably by controlling the mean of the entry distribution in the generated correlation matrices. This allows for the generation of correlation matrices that better represent realistic data and can be used to benchmark statistical methods for graph inference.}

\maketitle


\section{Introduction} \label{sec:intro}

Les modèles graphiques permettent de représenter les dépendances entre variables aléatoires. Ce domaine a suscité un grand intérêt ces dernières années, voir par exemple~\cite{kimmig2015lifted,  pearl2014probabilistic, koller2009probabilistic}, \cite[Chapitre 7]{Giraud} et les références associées. Le champ d'application est vaste, incluant la génétique~\cite{grechkin2015pathway}, l'étude des protéines~\cite{akbani2014pan}, la caractérisation des maladies~\cite{armour2017network}, la connectivité fonctionnelle du cerveau~\cite{huang2010learning} ou encore la gestion des risques~\cite{hull2012risk}. L'idée principale consiste à inférer une structure de graphe associée à la matrice de corrélation ou à la matrice de précision (inverse de la matrice de corrélation). Pour évaluer la qualité des procédures d'estimation, il est essentiel de pouvoir générer des matrices de corrélation ou de précision (théoriques) associées à une structure de graphe donnée, ce qui implique d'imposer des zéros particuliers dans la matrice tout en garantissant sa positivité, ce qui est généralement non trivial. L'objectif de cet article est de présenter une méthode permettant de générer de telles matrices.

Parmi les méthodes de génération de matrices de corrélation proposées dans la littérature, on peut citer \textit{vines} et \textit{onion}\cite{lewandowski2009generating}, basées sur la distribution Beta établie par \cite{joe2006generating}. Une autre approche consiste à utiliser la décomposition de Cholesky, comme dans \cite{pourahmadi2015distribution, cordoba2020generating}. Nous renvoyons à ces articles pour un aperçu bibliographique des méthodes existantes. Dans\cite{achard2022generation}, il a été observé que la distribution des corrélations de connectivité cérébrale était centrée autour de valeurs positives. Cependant, les méthodes existantes génèrent toutes des matrices de corrélation dont la distribution des entrées est centrée autour de zéro. Notre objectif est de proposer une nouvelle approche basée sur l'optimisation convexe permettant de contrôler cette distribution, en particulier sa moyenne.

Cet article\footnote{Ce travail a été soutenu par l'Agence Nationale de la Recherche dans le cadre du programme France 2030, référence ANR-23-IACL-0006.} est organisé comme suit. La Section~\ref{sec:notation} introduit les principales définitions et notations. La Section~\ref{sec:related} passe en revue les travaux connexes. La Section~\ref{sec:proposed} décrit l'approche proposée, et la Section~\ref{sec:results} présente les résultats ainsi qu'une comparaison avec d'autres approches.

\section{Notations} \label{sec:notation}

Une matrice réelle symétrique $\mathbf{A}$ de dimension $p \times p$ est semi-définie positive (SDP) si $\mathbfit{x}^\top \mathbf{A} \mathbfit{x} \geq 0$ pour tout $\mathbfit{x} \in \mathbb{R}^p$. Soit $\mathbfit{x} \in \mathbb{R}^{p}$ un vecteur aléatoire de matrice de covariance 
$
	\mathbf{\Sigma} = \mathbb{E} [(\mathbfit{x} - \mathbb{E}[\mathbfit x])(\mathbfit{x} - \mathbb{E}[\mathbfit x])^\top].
$
La matrice de corrélation associée $\cb \in \mathcal{C}$ est définie par 
$\cb = (\text{diag}(\mathbf \Sigma))^{-\frac{1}{2}} \ \mathbf \Sigma \ (\text{diag}(\mathbf \Sigma))^{-\frac{1}{2}} $.
L'ensemble $\mathcal{C}$ des matrices de corrélation satisfait :
\begin{equation} \label{eq:condition} 
\forall\cb \in \mathcal{C}, \; \text{diag}(\cb) = 1, \; \forall i, j \in {1, \dots, p}, \; -1\leq c_{ij} \leq 1.
\end{equation}
Générer une matrice de corrélation revient à construire une matrice symétrique SDP vérifiant~\eqref{eq:condition}~\cite[Problème 7.1.]{horn2012matrix}.
Dans le cas qui nous intéresse, nous cherchons une matrice de corrélation $\cb$ associée à un graphe $\gggg = (\vvv, \eee)$, c'est-à-dire vérifiant $c_{ij} = 0$ si $(i, j) \notin \eee$ et $c_{ij} \neq 0$ sinon, pour tout $(i,j)\in\vvv \times \vvv$ avec $i\neq j$. Les poids des arêtes dans le graphe correspondent aux valeurs de la matrice de corrélation. Nous définissons $\overline{\eee}$ comme l'ensemble des non-arêtes, correspondant aux entrées nulles de la matrice $\cb$. Ainsi, notre objectif est de générer une matrice de corrélation avec un ensemble prescrit $\overline{\eee}$ d'entrées nulles. Ce problème peut être vu comme un problème de complétion de matrice~\cite[Chapitre 10]{vandenberghe2015chordal}.

Dans la suite, nous notons $\cc(\gggg)$ l'ensemble des matrices de corrélation associées à un graphe donné $\gggg$ satisfaisant :
\begin{equation} \label{eq:constraint1} \cb = (c_{ij}) \text{ est SDP, satisfait \eqref{eq:condition}, et} \; c_{ij} = 0, \ (i, j) \notin \eee. \end{equation}
Nous considérons différentes structures de graphe, à savoir les graphes aléatoires d'\emph{Erdős-Rényi}, de \emph{Barabási-Albert}, et de \emph{Watts-Strogatz}, \cite{albert2002statistical}, ainsi que des \emph{Modèles à blocs stochastiques} \cite{abbe2018community} et des graphes \emph{cordaux} où tout cycle de longueur supérieure à trois possède une arête reliant deux sommets non adjacents dans le cycle. En pratique, nous générons un graphe cordal à partir d'un graphe de Barabási-Albert en ajoutant des arêtes si nécessaire pour satisfaire cette propriété.

Une caractéristique clé de la structure du graphe est sa densité, qui est le rapport entre le nombre d'arêtes et le nombre maximal d'arêtes possibles, $d = \frac{2|\eee|}{p(p-1)}$. 


\section{Travaux connexes} \label{sec:related}

Dans cette section, nous  présentons brièvement les méthodes qui, à notre connaissance, peuvent générer des matrices de corrélation associées à un graphe donné $\gggg$. La première approche repose sur la décomposition de Cholesky. Nous définissons $\mathcal{U}(\gggg) \subset \mathcal{U}$ comme le sous-ensemble des matrices triangulaires supérieures avec des éléments diagonaux positifs et lignes normalisées à 1, où $u_{ij} = 0$ pour tous les \mbox{$(i,j) \in \overline{\eee}$}. Si le graphe est cordal \cite[Chapitre 4]{vandenberghe2015chordal}, il est possible de générer le facteur de Cholesky $\mat{U}$ dans $\mathcal{U}(\gggg)$ en imposant $u_{ij} = 0, \forall (i, j) \in \overline{\eee}$. On a alors $\cb = \mat{U} \mat{U}^\top \in \cc(\gggg)$.

Dans~\cite{pourahmadi2015distribution}, les auteurs proposent une paramétrisation polaire des entrées du facteur de Cholesky $\mat{U} \in \mathcal{U}$ et établissent la distribution de probabilité à adopter de telle sorte que l’échantillonnage des matrices résultantes soit uniforme sur l'ensemble $\cc$. En utilisant cette paramétrisation polaire, il est facile d'incorporer la contrainte $u_{ij} = 0$ pour tous $(i,j) \in \overline{\eee}$ pour obtenir $\mat{U} \in \mathcal{U}(\gggg)$. Ceci assure que $\cb = \mat{U} \mat{U}^\top \in \cc(\gggg)$ mais pour les graphes cordaux uniquement. Dans~\cite{cordoba2020generating}, la méthode de génération proposée est basée sur l'algorithme Metropolis-Hastings~\cite{chib1995understanding} et donne alors une distribution uniforme sur $\cc(\gggg)$, mais n'est applicable là aussi que pour les graphes cordaux.


À notre connaissance, seules deux méthodes ont été proposées dans la littérature pour générer des matrices de corrélation associées à un graphe donné $\gggg$ sans exiger une structure cordale : la dominance diagonale et l'orthogonalisation partielle, toutes deux présentées dans \cite{cordoba2020generating}. Un inconvénient majeur de la dominance diagonale est qu'elle donne des matrices de corrélation avec des valeurs hors-diagonales très faibles. La méthode d'orthogonalisation partielle ne souffre pas de cette limitation, néanmoins notre méthode présentée à la Section \ref{sec:proposed} est moins sensible à la matrice initiale utilisée (en particulier parce que l'orthogonalisation partielle dépend de l'ordonnancement des nœuds) et permet de s'affranchir de la symétrie autour de zéro.

\section{Approche proposée} \label{sec:proposed}

L'objectif de notre travail est de générer des matrices de corrélation dans $\cc(\gggg)$, c'est-à-dire satisfaisant les contraintes \eqref{eq:constraint1}. Des contraintes supplémentaires peuvent être ajoutées, en fonction du contexte. L'une de nos motivations est de construire des matrices de corrélation avec une distribution qui ressemble aux données réelles, typiquement en neurosciences, où celles-ci sont décalées vers des valeurs positives \cite{achard2022generation}. Pour refléter cette propriété, nous imposons cette contrainte supplémentaire sur la moyenne : pour $b \geq -1$, \begin{equation}\label{eq:b} \frac{1}{2\vert \eee \vert}\sum_{i\neq j} c_{ij} \geq b. \end{equation} Prendre $b \leq -1$ revient à ne pas imposer de contrainte.

Nous cherchons à résoudre le problème d'optimisation : 
\begin{equation} \label{eq:constraint2} 
\begin{aligned} 
& \underset{\cb}{\text{minimiser}} & & \frac{1}{2}\lVert \cb - \bar{\cb} \rVert^2_F, & & \text{sous contraintes } \eqref{eq:constraint1} \text{ et } \eqref{eq:b}, 
\end{aligned} 
\end{equation} 
où $\bar{\cb}$ est une matrice donnée arbitraire. Avec des données réelles, cela peut être la matrice de corrélation empirique. Il convient de noter que la résolution de \eqref{eq:constraint2} garantit que la moyenne des éléments non diagonaux est au moins égale à $b$. Puisque la fonction objectif dans \eqref{eq:constraint2} est convexe, une solution existe chaque fois que les contraintes sont réalisables. Par ailleurs, la matrice identité satisfait les contraintes \eqref{eq:constraint1}, garantissant ainsi la faisabilité en l'absence de la contrainte supplémentaire \eqref{eq:b}. Dans la Section \ref{sec:results}, nous examinons l'impact de cette contrainte supplémentaire sur l'existence de solutions.

\section{Résultats et discussions} \label{sec:results}

Dans nos simulations, nous considérons $\cb \in \mathbb{R}^{51\times51}$ et $\bar \cb$ une matrice de même taille dont les entrées suivent une loi uniforme sur l'intervalle $[-1, 1]$. Pour le motif $\eee$, nous utilisons les différents modèles de graphes aléatoires mentionnés à la Section \ref{sec:notation}. Le problème d'optimisation \eqref{eq:constraint2} est résolu en utilisant le solveur \textsc{cvxopt} de la bibliothèque Python \textsc{CVXPY}~\cite{cvxpy} qui implémente une méthode de point intérieur. Nous l'avons appliqué à ces modèles de graphes sur 50 exécutions\footnote{Les simulations ont été réalisées à l'aide de l'infrastructure GRICAD, soutenue par la communauté de recherche grenobloise. Le code pour reproduire les expériences est disponible en ligne \cite{code}.}. Dans certains cas, la résolution de \eqref{eq:constraint2} aboutit numériquement à une matrice dont la valeur propre minimale est proche de zéro, mais négative, ce qui indique que la matrice n'est pas strictement SDP. Pour y remédier, nous considérons la matrice $\tilde{\mathbf{C}}_{\epsilon} = \tilde{\mathbf{C}} + \epsilon \mathbf{I}$ (avec $\epsilon=10^{-8}$) que nous renormalisons par $\mathbf{C} = \frac{1}{1+\epsilon}\tilde{\mathbf{C}}_{\epsilon}$. La matrice $\mathbf{C}$ n'est pas le minimiseur de la fonction objectif, mais elle est une matrice de corrélation qui satisfait les contraintes \eqref{eq:constraint2}\footnote{Pour être plus précis, avec cette étape de post-traitement, la valeur moyenne change, et donc la contrainte \eqref{eq:b} peut ne pas être satisfaite. Augmenter $b$ à $b(1+\epsilon)$ permet d'atteindre notre objectif.}.

\subsection{Comparaison avec d'autres approches}

\begin{figure}[htbp] 
\centerline{\includegraphics[width=0.4\textwidth]{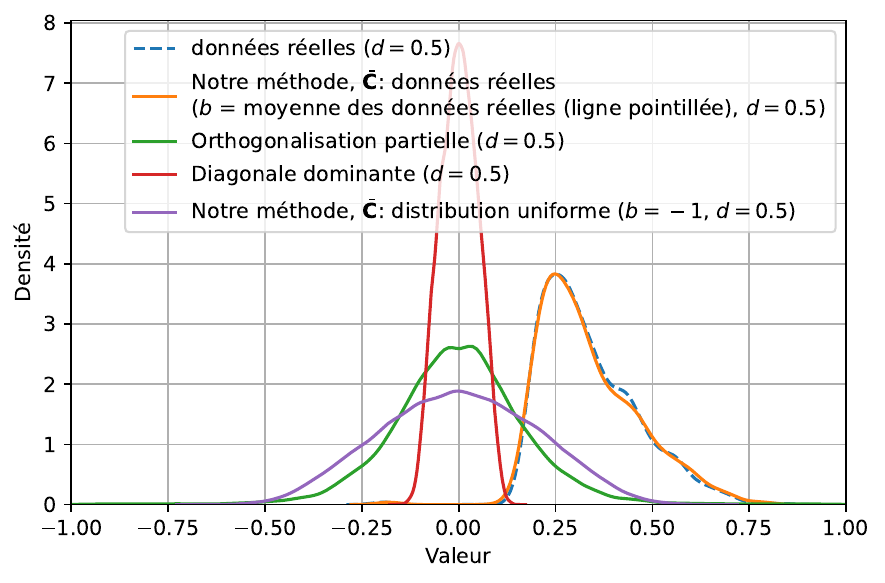}} 
\vspace*{-0.5\baselineskip}
\caption{Densité des éléments non diagonaux non nuls dans les matrices de corrélation générées par différentes méthodes (dominance diagonale, orthogonalisation partielle et notre méthode) comparées à la matrice de corrélation obtenue à partir des données fMRI de rats.} \label{fig:correlations3} 
\end{figure}

Pour la comparaison, nous considérons un graphe avec 51 nœuds, c'est-à-dire $\mathbf{C} \in \mathbb{R}^{51 \times 51}$. La Figure~\ref{fig:correlations3} montre la densité des éléments non diagonaux non nuls. Nous comparons notre méthode avec deux autres approches : la dominance diagonale et l'orthogonalisation partielle. Plus précisément, nous générons 50 graphes d'\emph{Erdős-Rényi} en utilisant la dominance diagonale, l'orthogonalisation partielle, et la méthode proposée. Pour cette dernière, nous fixons la densité cible du graphe à 0.5. La distribution des valeurs de corrélation est représentée par les lignes rouge, verte et violette, respectivement - la ligne orange est liée aux données réelles et expliquée ci-dessous. Nous fixons le paramètre $b = -1$ dans notre algorithme pour faciliter la comparaison avec d'autres algorithmes qui n'utilisent pas de seuil. Dans notre algorithme, à la fois $\bar{\mathbf{C}}$ et le point initial pour la méthode de dominance diagonale sont des réalisations d'une distribution uniforme, puisque nous visons à générer des matrices de corrélation aléatoires. Il convient de noter que, lorsqu'on utilise la dominance diagonale, aucune perturbation positive n'est appliquée pour garantir que la matrice est SDP. Comme mentionné précédemment, les densités obtenues avec la dominance diagonale sont concentrées autour de faibles valeurs (ligne rouge). Notre approche (ligne violette) donne des entrées plus élevées dans la matrice de corrélation. Nous disposons ainsi d'un modèle de génération de matrices de corrélations synthétiques à graphe fixé, dont les valeurs de corrélations significatives ne se confondent pas avec le bruit, nous permettant ainsi dans \cite{chevauxBenchmarking2025} d'effectuer un \textit{benchmark} de différentes techniques d'inférence de graphes.

\subsection{Influence de la contrainte sur la moyenne}

Dans la section précédente, nous avons montré que la méthode proposée génère des corrélations non négligeables et dont la distribution est symétrique autour de zéro. En revanche dans certains jeux de données réelles, la distribution des corrélations empiriques est décalée vers les valeurs positives. Or à notre connaissance peu de travaux ont concerné la simulation de matrices de corrélation semblables aux données réelles, à l'exception de \cite{marti} utilisant des réseaux antagonistes génératifs (GANs), mais qui nécessitent un grand ensemble de données de matrices de corrélation observées, ce qui peut ne pas toujours être disponible en pratique.
Nous examinons maintenant l'impact de la contrainte~\eqref{eq:b}, qui modifie le centre de la distribution des valeurs de corrélation. En particulier, cela permet de générer des matrices de corrélation plus proches des données réelles et cela peut être utile pour contrôler le rapport signal/bruit dans les simulations.

Dans notre contexte, nous sommes motivés par une application en neurosciences impliquant des données d'IRM fonctionnelle acquises sur des rats. Les données sont décrites et en accès libre~\cite{guillaume, achard2022generation}. Dans la Figure~\ref{fig:correlations3}, la ligne bleue représente la distribution des corrélation (empirique) des données IRM d'un rat, tandis que la ligne orange montre la distribution des entrées de la matrice de corrélation (théorique) générée à l'aide de notre méthode. Pour les données réelles, nous calculons la densité du graphe en sélectionnant les 50\% des entrées avec les valeurs absolues les plus élevées dans la matrice de corrélation. Lors de la génération de la matrice synthétique avec notre méthode, nous fixons $d = 50\%$ et le paramètre $b$ égal à la moyenne des entrées de la matrice de corrélation des données réelles correspondant au graphe calculé~($d = 50\%$).  La matrice initiale $\bar{\cb}$ est ici égale à la corrélation empirique des données réelles. La Figure~\ref{fig:correlations3} montre que la distribution des données simulées non nulles est en effet proche de celle des données réelles décalées vers le positif, avec en sus la contrainte des zéros satisfaite, ce qui n’aurait pas été réalisable avec les méthodes présentées à la Section \ref{sec:related}. À noter qu’un choix aléatoire pour $\bar{\cb}$ eut été également possible pour approcher cette distribution une fois la moyenne $b$ adéquatement fixée, nous avons choisi ici d’illustrer la projection de la matrices de corrélation empirique sur les contraintes de faisabilité. 

\subsection{Influence de la structure du graphe} 

\begin{figure*}
\subfloat[\emph{Erdős-Rényi}]{ \includegraphics[width=0.3\textwidth]{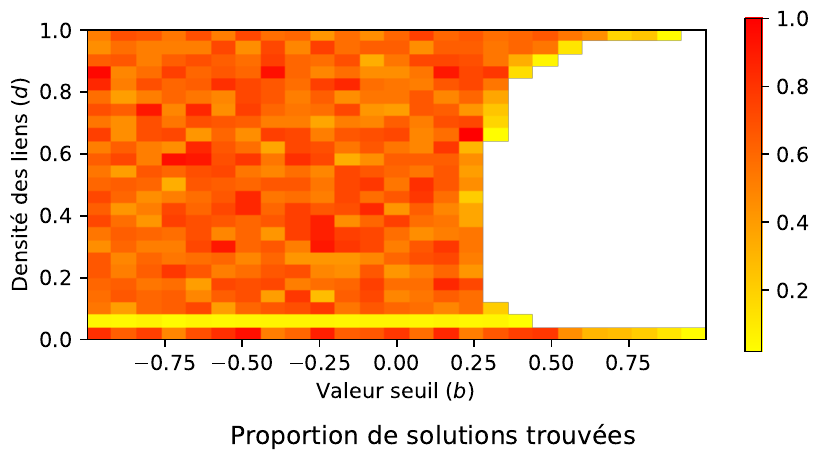}\label{figa}
    }
\subfloat[\emph{cordaux}]{ \includegraphics[width=0.3\textwidth]{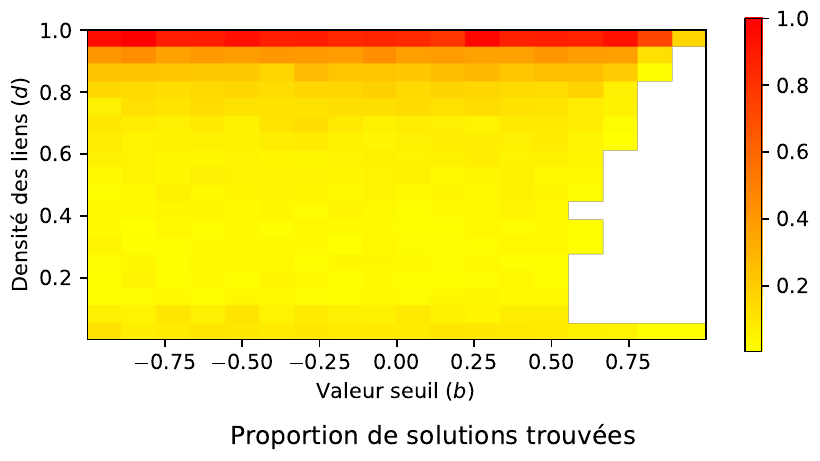}\label{figb}
    }
\subfloat[\emph{Watts–Strogatz}]{ \includegraphics[width=0.3\textwidth]{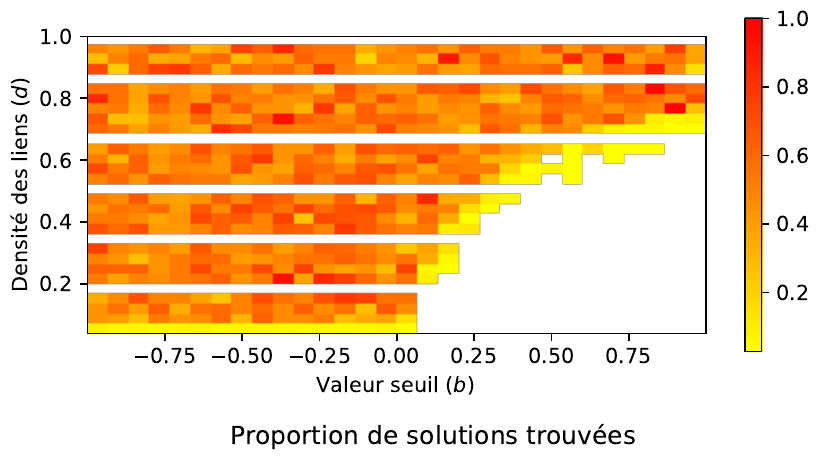}\label{figc}
    }
\caption{Proportion de réussites pour trouver une matrice de corrélation $\cb \in \mathbb{R}^{51 \times 51}$ par bin en fonction des paramètres $(b,d)$, utilisant les modèles (a) \emph{Erdős-Rényi}, (b) \emph{cordaux} et (c) \emph{Watts–Strogatz}. Les zones blanches correspondent à une absence de solution.} 
\label{fig:solution} 
\end{figure*}

L'ajout de la contrainte~\eqref{eq:b} peut parfois aboutir à un problème d'optimisation sans solution. La Figure~\ref{fig:solution} illustre la proportion de cas où une matrice de corrélation valide \mbox{$\cb \in \mathbb{R}^{51 \times 51}$} est trouvée pour différentes densités de graphes $d$ et valeurs de moyenne $b$ dans~\eqref{eq:constraint2}, et différentes structures de graphes. 

\begin{figure}[!ht]
\vspace{-0.5cm}
\includegraphics[width=0.4\textwidth]{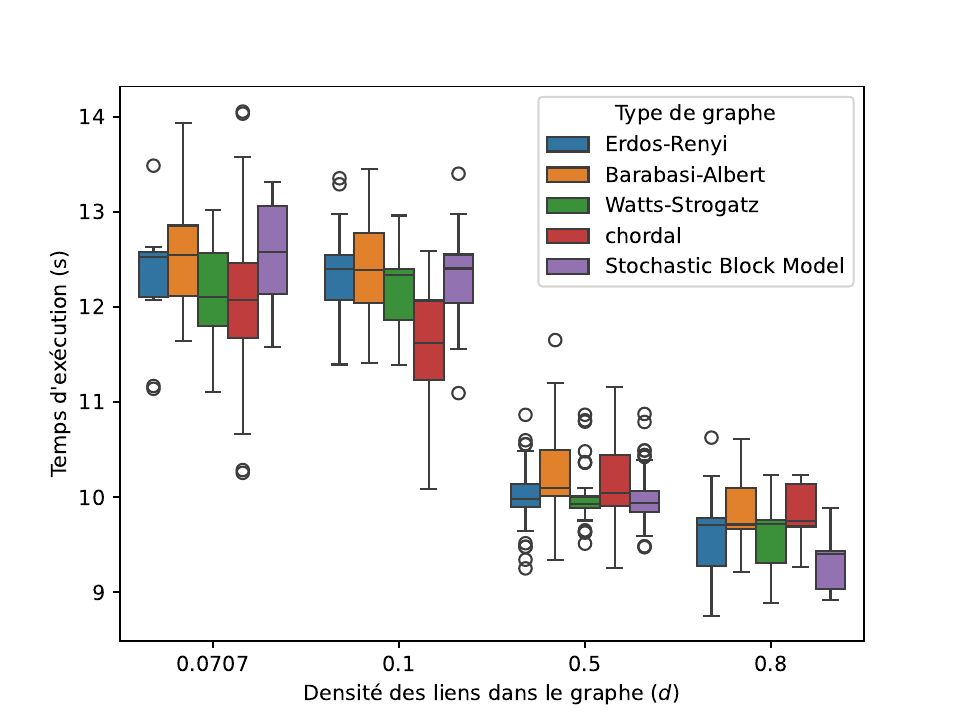}
\caption{Temps d'exécution (en secondes) de notre méthode pour calculer la matrice de corrélation $\cb \in \mathbb{R}^{51 \times 51}$, moyenné sur 50 exécutions pour chaque boîte à moustaches (représentant un type de graphe différent) en fonction de la densité des éléments non nuls et non diagonaux dans la matrice de corrélation. Le paramètre $b$ est fixé à~$-1$.}
\label{fig:time}
\end{figure}

Le coût computationnel de la résolution du problème d'optimisation est significativement plus élevé que pour des méthodes comme la dominance diagonale ou l'orthogonalisation partielle qui n'excèdent pas la dizaine de secondes. L'augmentation de la dimension $p$ de $\mathbf{C}$ entraîne généralement une augmentation du temps d'exécution. La Figure~\ref{fig:time} compare les temps d'exécution pour $p=51$ avec différentes densités de graphes et modèles. Globalement, le temps d'exécution diminue à mesure que la densité $d$ du graphe augmente. Seul le cas $b=-1$ est
représenté, mais des résultats similaires sont obtenus pour différentes valeurs de $b$.

\section*{Conclusion et perspectives}

La méthode proposée pour générer une matrice de corrélation correspondant à un graphe présente plusieurs avantages: elle ne repose pas sur une structure de type cordale, elle évite de générer des valeurs trop faibles en contrôlant la moyenne, et enfin elle peut approcher une matrice de corrélation empirique. Cependant, le coût de cette approche augmente avec la dimension. Puisque nous projetons à court terme d'étudier comment l'augmentation du nombre de nœuds affecte l'influence des structures de graphes, l'utilisation de l'algorithme QSDPNAL~\cite{li2018qsdpnal} en MATLAB pourrait être envisagé pour résoudre le problème d'optimisation en plus grande dimension. Enfin une perspective consistera à étudier l'échantillonnage de $\bar{\cb}$ pouvant conduire à une distribution uniforme sur $\cc(\gggg)$.

{\small 
\bibliography{biblio}

\begin{thebibliography}{10}
\expandafter\ifx\csname fonteauteurs\endcsname\relax
\def\fonteauteurs{\scshape}\fi

\bibitem{abbe2018community}
Emmanuel \bgroup\fonteauteurs\bgroup Abbe\egroup\egroup{} :
\newblock Community detection and stochastic block models: recent developments.
\newblock {\em Journal of Machine Learning Research},
  18(177)\string:\penalty500\relax 1--86, 2018.

\bibitem{achard2022generation}
Sophie \bgroup\fonteauteurs\bgroup Achard\egroup\egroup{}, Ir{\`e}ne
  \bgroup\fonteauteurs\bgroup Gannaz\egroup\egroup{} et K{\'e}vin
  \bgroup\fonteauteurs\bgroup Polisano\egroup\egroup{} :
\newblock G{\'e}n{\'e}ration de mod{\`e}les graphiques.
\newblock \emph{In} {\em GRETSI 2022-XXVIII{\`e}me Colloque francophone de
  traitement du signal et des images}, pages 1--3, 2022.

\bibitem{akbani2014pan}
Rehan \bgroup\fonteauteurs\bgroup Akbani\egroup\egroup{}, Patrick Kwok~Shing
  \bgroup\fonteauteurs\bgroup Ng\egroup\egroup{}, Henrica~MJ
  \bgroup\fonteauteurs\bgroup Werner\egroup\egroup{}, Maria
  \bgroup\fonteauteurs\bgroup Shahmoradgoli\egroup\egroup{}, Fan
  \bgroup\fonteauteurs\bgroup Zhang\egroup\egroup{}, Zhenlin
  \bgroup\fonteauteurs\bgroup Ju\egroup\egroup{}, Wenbin
  \bgroup\fonteauteurs\bgroup Liu\egroup\egroup{}, Ji-Yeon
  \bgroup\fonteauteurs\bgroup Yang\egroup\egroup{}, Kosuke
  \bgroup\fonteauteurs\bgroup Yoshihara\egroup\egroup{}, Jun
  \bgroup\fonteauteurs\bgroup Li\egroup\egroup{} \emph{et~al.} :
\newblock A pan-cancer proteomic perspective on the {C}ancer {G}enome {A}tlas.
\newblock {\em Nature communications}, 5(1)\string:\penalty500\relax 3887,
  2014.

\bibitem{albert2002statistical}
R{\'e}ka \bgroup\fonteauteurs\bgroup Albert\egroup\egroup{} et
  Albert-L{\'a}szl{\'o} \bgroup\fonteauteurs\bgroup
  Barab{\'a}si\egroup\egroup{} :
\newblock Statistical mechanics of complex networks.
\newblock {\em Reviews of modern physics}, 74(1)\string:\penalty500\relax 47,
  2002.

\bibitem{armour2017network}
Cherie \bgroup\fonteauteurs\bgroup Armour\egroup\egroup{}, Eiko~I
  \bgroup\fonteauteurs\bgroup Fried\egroup\egroup{}, Marie~K
  \bgroup\fonteauteurs\bgroup Deserno\egroup\egroup{}, Jack
  \bgroup\fonteauteurs\bgroup Tsai\egroup\egroup{} et Robert~H
  \bgroup\fonteauteurs\bgroup Pietrzak\egroup\egroup{} :
\newblock A network analysis of {DSM-5} posttraumatic stress disorder symptoms
  and correlates in {US} military veterans.
\newblock {\em Journal of anxiety disorders}, 45\string:\penalty500\relax
  49--59, 2017.

\bibitem{guillaume}
Guillaume J-PC \bgroup\fonteauteurs\bgroup Becq\egroup\egroup{}, Tarik
  \bgroup\fonteauteurs\bgroup Habet\egroup\egroup{}, Nora
  \bgroup\fonteauteurs\bgroup Collomb\egroup\egroup{}, Margaux
  \bgroup\fonteauteurs\bgroup Faucher\egroup\egroup{}, Chantal
  \bgroup\fonteauteurs\bgroup Delon-Martin\egroup\egroup{}, V{\'e}ronique
  \bgroup\fonteauteurs\bgroup Coizet\egroup\egroup{}, Sophie
  \bgroup\fonteauteurs\bgroup Achard\egroup\egroup{} et Emmanuel~L
  \bgroup\fonteauteurs\bgroup Barbier\egroup\egroup{} :
\newblock Functional connectivity is preserved but reorganized across several
  anesthetic regimes.
\newblock {\em NeuroImage}, 219\string:\penalty500\relax 116945, 2020.

\bibitem{chevauxBenchmarking2025}
Alice \bgroup\fonteauteurs\bgroup Chevaux\egroup\egroup{}, Ali
  \bgroup\fonteauteurs\bgroup Fahkar\egroup\egroup{}, K{\'e}vin
  \bgroup\fonteauteurs\bgroup Polisano\egroup\egroup{}, Ir{\`e}ne
  \bgroup\fonteauteurs\bgroup Gannaz\egroup\egroup{} et Sophie
  \bgroup\fonteauteurs\bgroup Achard\egroup\egroup{} :
\newblock {Benchmarking Brain Connectivity Graph Inference: A Novel Validation
  Approach}.
\newblock \emph{In} {\em {33rd European Signal Processing Conference (EUSIPCO
  2025)}}, Palerme, Italy, septembre 2025.

\bibitem{chib1995understanding}
Siddhartha \bgroup\fonteauteurs\bgroup Chib\egroup\egroup{} et Edward
  \bgroup\fonteauteurs\bgroup Greenberg\egroup\egroup{} :
\newblock Understanding the {M}etropolis-{H}astings algorithm.
\newblock {\em The American Statistician}, 49(4)\string:\penalty500\relax
  327--335, 1995.

\bibitem{cordoba2020generating}
Irene \bgroup\fonteauteurs\bgroup C{\'o}rdoba\egroup\egroup{}, Gherardo
  \bgroup\fonteauteurs\bgroup Varando\egroup\egroup{}, Concha
  \bgroup\fonteauteurs\bgroup Bielza\egroup\egroup{} et Pedro
  \bgroup\fonteauteurs\bgroup Larra{\~n}aga\egroup\egroup{} :
\newblock On generating random {G}aussian graphical models.
\newblock {\em International Journal of Approximate Reasoning},
  125\string:\penalty500\relax 240--250, 2020.

\bibitem{cvxpy}
Steven \bgroup\fonteauteurs\bgroup Diamond\egroup\egroup{} et Stephen
  \bgroup\fonteauteurs\bgroup Boyd\egroup\egroup{} :
\newblock {CVXPY}: A {P}ython-embedded modeling language for convex
  optimization.
\newblock {\em Journal of Machine Learning Research},
  17(83)\string:\penalty500\relax 1--5, 2016.

\bibitem{code}
Ali \bgroup\fonteauteurs\bgroup Fahkar\egroup\egroup{}, Polisano
  \bgroup\fonteauteurs\bgroup Kévin\egroup\egroup{}, Irène
  \bgroup\fonteauteurs\bgroup Gannaz\egroup\egroup{} et Sophie
  \bgroup\fonteauteurs\bgroup Achard\egroup\egroup{} :
\newblock Code of the present paper, 2025.
\newblock
  https://gricad-gitlab.univ-grenoble-alpes.fr/polisank/generating-correlation-matrices-with-graph-structures-using-convex-optimization.

\bibitem{Giraud}
Christophe \bgroup\fonteauteurs\bgroup Giraud\egroup\egroup{} :
\newblock {\em Introduction to high-dimensional statistics}.
\newblock CRC Press, 2021.

\bibitem{grechkin2015pathway}
Maxim \bgroup\fonteauteurs\bgroup Grechkin\egroup\egroup{}, Maryam
  \bgroup\fonteauteurs\bgroup Fazel\egroup\egroup{}, Daniela
  \bgroup\fonteauteurs\bgroup Witten\egroup\egroup{} et Su-In
  \bgroup\fonteauteurs\bgroup Lee\egroup\egroup{} :
\newblock Pathway graphical lasso.
\newblock {\em Proceedings of the AAAI conference on artificial intelligence},
  29(1), 2015.

\bibitem{horn2012matrix}
Roger~A \bgroup\fonteauteurs\bgroup Horn\egroup\egroup{} et Charles~R
  \bgroup\fonteauteurs\bgroup Johnson\egroup\egroup{} :
\newblock {\em Matrix analysis}.
\newblock Cambridge University Press, 2012.

\bibitem{huang2010learning}
Shuai \bgroup\fonteauteurs\bgroup Huang\egroup\egroup{}, Jing
  \bgroup\fonteauteurs\bgroup Li\egroup\egroup{}, Liang
  \bgroup\fonteauteurs\bgroup Sun\egroup\egroup{}, Jieping
  \bgroup\fonteauteurs\bgroup Ye\egroup\egroup{}, Adam
  \bgroup\fonteauteurs\bgroup Fleisher\egroup\egroup{}, Teresa
  \bgroup\fonteauteurs\bgroup Wu\egroup\egroup{}, Kewei
  \bgroup\fonteauteurs\bgroup Chen\egroup\egroup{}, Eric
  \bgroup\fonteauteurs\bgroup Reiman\egroup\egroup{} et Alzheimer's
  Disease~NeuroImaging \bgroup\fonteauteurs\bgroup Initiative\egroup\egroup{} :
\newblock Learning brain connectivity of {A}lzheimer's disease by sparse
  inverse covariance estimation.
\newblock {\em NeuroImage}, 50(3)\string:\penalty500\relax 935--949, 2010.

\bibitem{hull2012risk}
John \bgroup\fonteauteurs\bgroup Hull\egroup\egroup{} :
\newblock {\em Risk management and financial institutions,+ Web Site}, volume
  733.
\newblock John Wiley \& Sons, 2012.

\bibitem{joe2006generating}
Harry \bgroup\fonteauteurs\bgroup Joe\egroup\egroup{} :
\newblock Generating random correlation matrices based on partial correlations.
\newblock {\em Journal of Multivariate Analysis},
  97(10)\string:\penalty500\relax 2177--2189, 2006.

\bibitem{kimmig2015lifted}
Angelika \bgroup\fonteauteurs\bgroup Kimmig\egroup\egroup{}, Lilyana
  \bgroup\fonteauteurs\bgroup Mihalkova\egroup\egroup{} et Lise
  \bgroup\fonteauteurs\bgroup Getoor\egroup\egroup{} :
\newblock Lifted graphical models: a survey.
\newblock {\em Machine Learning}, 99\string:\penalty500\relax 1--45, 2015.

\bibitem{koller2009probabilistic}
Daphne \bgroup\fonteauteurs\bgroup Koller\egroup\egroup{} et Nir
  \bgroup\fonteauteurs\bgroup Friedman\egroup\egroup{} :
\newblock {\em Probabilistic graphical models: principles and techniques}.
\newblock MIT press, 2009.

\bibitem{lewandowski2009generating}
Daniel \bgroup\fonteauteurs\bgroup Lewandowski\egroup\egroup{}, Dorota
  \bgroup\fonteauteurs\bgroup Kurowicka\egroup\egroup{} et Harry
  \bgroup\fonteauteurs\bgroup Joe\egroup\egroup{} :
\newblock Generating random correlation matrices based on vines and extended
  onion method.
\newblock {\em Journal of multivariate analysis},
  100(9)\string:\penalty500\relax 1989--2001, 2009.

\bibitem{li2018qsdpnal}
Xudong \bgroup\fonteauteurs\bgroup Li\egroup\egroup{}, Defeng
  \bgroup\fonteauteurs\bgroup Sun\egroup\egroup{} et Kim-Chuan
  \bgroup\fonteauteurs\bgroup Toh\egroup\egroup{} :
\newblock {QSDPNAL}: A two-phase augmented lagrangian method for convex
  quadratic semidefinite programming.
\newblock {\em Mathematical Programming Computation},
  10\string:\penalty500\relax 703--743, 2018.

\bibitem{marti}
Gautier \bgroup\fonteauteurs\bgroup Marti\egroup\egroup{}, Victor
  \bgroup\fonteauteurs\bgroup Goubet\egroup\egroup{} et Frank
  \bgroup\fonteauteurs\bgroup Nielsen\egroup\egroup{} :
\newblock {cCorrGAN}: Conditional correlation gan for learning empirical
  conditional distributions in the elliptope.
\newblock \emph{In} {\em International Conference on Geometric Science of
  Information}, pages 613--620. Springer, 2021.

\bibitem{pearl2014probabilistic}
Judea \bgroup\fonteauteurs\bgroup Pearl\egroup\egroup{} :
\newblock {\em Probabilistic reasoning in intelligent systems: networks of
  plausible inference}.
\newblock Elsevier, 2014.

\bibitem{pourahmadi2015distribution}
Mohsen \bgroup\fonteauteurs\bgroup Pourahmadi\egroup\egroup{} et Xiao
  \bgroup\fonteauteurs\bgroup Wang\egroup\egroup{} :
\newblock Distribution of random correlation matrices: {H}yperspherical
  parameterization of the {C}holesky factor.
\newblock {\em Statistics \& Probability Letters}, 106\string:\penalty500\relax
  5--12, 2015.

\bibitem{vandenberghe2015chordal}
Lieven \bgroup\fonteauteurs\bgroup Vandenberghe\egroup\egroup{} et Martin~S
  \bgroup\fonteauteurs\bgroup Andersen\egroup\egroup{} :
\newblock Chordal graphs and semidefinite optimization.
\newblock {\em Foundations and Trends{\textregistered} in Optimization},
  1(4)\string:\penalty500\relax 241--433, 2015.

\end{thebibliography}
}


\end{document}